\pdfoutput=1
\documentclass[aps,prl,reprint,superscriptaddress,longbibliography,nofootinbib]{revtex4-1}

\usepackage{graphicx}
\usepackage{amsmath}
\usepackage{amssymb}
\usepackage{amsfonts}
\usepackage{dcolumn}
\usepackage{dsfont}
\usepackage{latexsym}
\usepackage{rotating}
\usepackage{color}
\usepackage{latexsym}
\usepackage{bbm}
\usepackage{subfigure}
\usepackage{float}
\usepackage{epsfig}
\usepackage{psfrag}
\usepackage{natbib, hyperref}
\usepackage{bm}
\usepackage{amsthm}
\usepackage{eucal}
\usepackage{mathrsfs}
\usepackage{url}
\usepackage{braket}
\usepackage{ulem}
\usepackage{calligra}
\usepackage[T1]{fontenc}
\usepackage[utf8]{inputenc}
\usepackage{newunicodechar}
\usepackage{marvosym}
\usepackage[absolute]{textpos}
\usepackage[cal=cm]{mathalfa}
\usepackage{soul}

\usepackage{color}

\usepackage{hyperref}
\hypersetup{
colorlinks=true,final=true,
        linkcolor=black,
        citecolor=black,
        filecolor=black,
        urlcolor=black,
}

\begin{document}
\setlength{\abovedisplayskip}{3pt}
\setlength{\belowdisplayskip}{3pt}

\title{Chiral pair density wave as a precursor of the pseudogap in kagom\'e superconductors}

\author{Narayan Mohanta\\
\textit{\small Department of Physics, Indian Institute of Technology Roorkee, Roorkee 247667, India}}

\begin{abstract}
Motivated by scanning tunneling microscopy experiments on $A$V$_3$Sb$_5$ ($A$ \!=\! Cs, Rb, K) that revealed periodic real-space modulation of electronic states at low energies, I show using model calculations that a triple-{\bf Q} chiral pair density wave (CPDW) is generated in the superconducting state by a charge order of $2a\! \times \!2a$ superlattice periodicity, intertwined with a time-reversal symmetry breaking orbital loop current. In the presence of such a charge order and orbital loop current, the superconducting critical field is enhanced beyond the Chandrasekhar-Clogston limit. The CPDW correlation survives even when the long-range superconducting phase coherence is diminished by a magnetic field or temperature, stabilizing an exotic granular superconducting state above and in the vicinity of the superconducting transition. The presented results suggest that the CPDW can be regarded as the origin of the pseudogap observed near the superconducting transition. 
\end{abstract}
           
\maketitle

%\textit{Introduction.}---
Understanding electronic properties arising from coexisting superconductivity and various density-wave orders has remained as a central problem in condensed matter physics. It has captivated the physics community for decades in the context of high-temperature cuprate superconductors; the recently-synthesized kagom\'e metals $A$V$_3$Sb$_5$ ($A$ \!=\! Cs, Rb, K) have revived the interest~\cite{PhysRevMaterials.3.094407}. The V atoms in these compounds form a kagom\'e lattice, and the Fermi level is predominantly populated by V $3d$ orbitals. The electronic band structure exhibits Dirac points and nearly-flat less-dispersive bands~\cite{PhysRevLett.125.247002,Shi_NComm2022}. Strong correlation of the nearly-flat bands, topological effects from the Dirac fermions, van Hove singularities, and frustration effects in the kagom\'e geometry are favorable conditions for instabilities towards long-range many-body order to set in. Superconductivity with a gap-to-$T_c$ ratio $2\Delta_0/k_{B}T_c\! \approx \!5$ was found below $T_c\! \approx \!2.5$~K~\cite{PhysRevLett.125.247002}. A chiral charge order was found to appear below $T_{\small {\rm co}}\! \approx \!94$~K with broken time-reversal symmetry (TRS) but without the trace of any long-range magnetic order, indicating the presence of an intertwined orbital loop current~\cite{Jiang2021,Mielke2022,Khasanov_PRRes2022}. The absence of acoustic phonon anomaly at the charge-order wave vector rules out the Pierls instability related to the Fermi surface nesting and phonon softening as a possible mechanism, and implies that extended Coulomb interactions at a van Hove filling may be responsible for it~\cite{PhysRevX.11.031050,Denner_PRL2021,Christensen_PRB2021}. A pressure-driven transition from fully-gapped to partially-gapped superconductivity and the coexistence of the superconductivity with the charge order over a large parameter regime suggest unconventional pairing in these compounds~\cite{Guguchia2023}. Alternative scenarios include non-chiral, anisotropic $s$-wave superconductivity, supported by recent experimental findings at different pressures~\cite{Roppongi2023}.

A suppressed electronic density of states at the Fermi level, known as the `pseudogap', posed an enigmatic problem in the high-temperature cuprate superconductors. A similar pseudogap with a V-shaped density of states was observed in scanning tunneling microscopy experiments on $A$V$_3$Sb$_5$ and subsequent theoretical analysis, with periodic modulations of both charge density and Cooper pair density of $2a\!\times \!2a$ superlattice periodicity ($a$ being the lattice constant)~\cite{Zhao2021,Chen2021,Xu_PRl2021,Jiang_PRB2023}. The findings are usually indicative of a nodal pairing symmetry or ungapped sections of the Fermi surface. The concomitant periodic modulations of both superfluid and normal fluid raised a series of questions including the origin of the pseudogap found in the tunneling spectra.

Here, I focus on the observed variation of the density of states in the pseudogap near the superconducting transition and show that a chiral density wave of $s$-wave Cooper pairs can account for it. The chiral pair density wave (CPDW) is generated in the superconducting state by the TRS breaking charge order and it persists above the superconducting transition without long-range superconducting phase coherence. This CPDW state can be described by a pairing gap $\Delta(\mathbf{r})\!=\!\sum_{a}\Delta_a e^{i(\mathbf{Q}_{a}\cdot \mathbf{r}+\varphi_a)}$ at a lattice site position $\mathbf{r}$, $\Delta_a$ and $\varphi_a$ being the magnitude and the relative phase of the pairing amplitude along three characteristic momentum $\mathbf{Q}_{a}$ ($a\!=\!1,2,3$), set by the charge order periodicity. The presented theoretical arguments are based on calculated density of states $\rho(E)$ and Fourier transformed local density of states $\rho({\mathbf Q_p}, E)$ at a CPDW wave vector ${\mathbf Q_p}$. In the vicinity and above the critical field $B_c$ or critical temperature $T_c$ for the superconducting transition, determined by a vanishing superfluid density $n_s$, $\rho({\mathbf Q_p}, E)$ reveals a particle-hole symmetric density of states around zero energy, thereby ruling out the charge-ordered  electronic states as a possible origin of the pseudogap. Remarkably, the critical value of the magnetic field, perpendicular to the kagom\'e plane, is found to be enhanced beyond the usual Chandrasekhar-Clogston limit in the presence of the orbital loop current, implying that a highly-provoking quantum state prevails above the superconducting transition.

Two types of charge order were reported---star of David and tri-hexagonal or inverse star of David patterns, both of $2a\! \times \!2a$ periodicity~\cite{Kato2022,Gupta2022,Tan_PRL2021}. A chiral flux phase, compatible with the symmetry of the kagom\'e lattice and broken TRS, was shown to be energetically favorable~\cite{FENG20211384}. The tri-hexagonal charge order which has been observed prominently in most compounds, as shown in Fig.~\ref{fig1}, is considered in the theoretical model and results presented below. Such an unusual charge ordered state is also supported by a number of interesting phenomena such as anomalous Hall effect and Nernst effect~\cite{Yang_sciadv2020,Gan_PRB2021,PhysRevB.104.L041103}.
%---------------------------------------------
\begin{figure}[t]
\begin{center}
\vspace{-0mm}
\epsfig{file=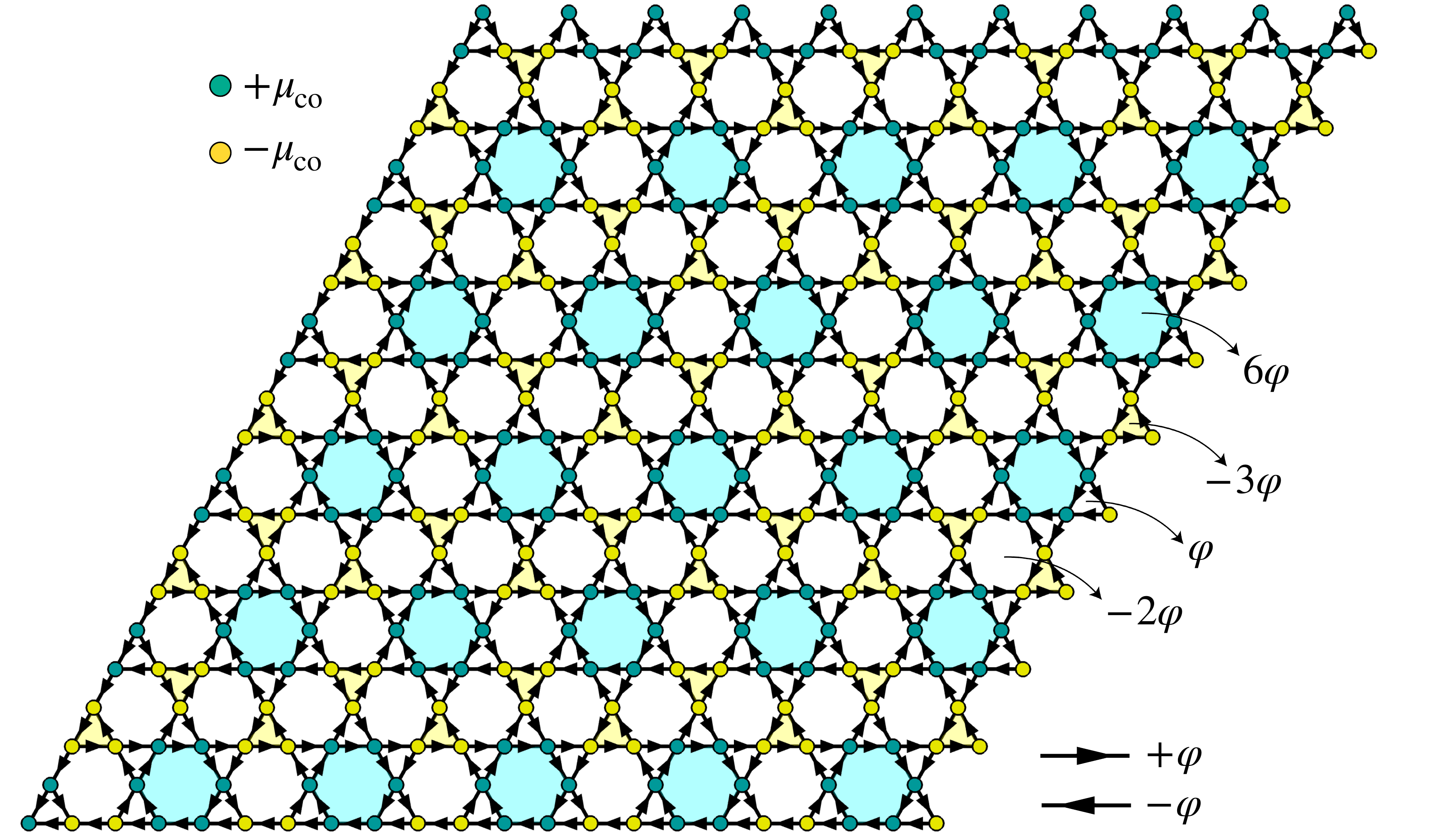,trim=0.0in 0.0in 0.0in 0.0in,clip=false, width=86mm}
\caption{Charge order configuration with intertwined orbital loop current on the kagom\'e lattice, analogous to the tri-hexagonal pattern observed in experiments. The yellow and cyan colors represent the modulation in the chemical potential $\mu_{\rm co}$, while the arrows represent the loop current propagation direction and the associated flux $\varphi$.}
\label{fig1}
\vspace{-4mm}
\end{center}
\end{figure}
%---------------------------------------------

\textit{Model and observables.}---To model the superconducting state and the experimentally-observed pseudogap, the minimal tight-binding Hamiltonian at the mean spin-singlet pairing field on the kagom\'e lattice is expressed as
\begin{align}
{\cal H}\!=&\!-t\sum_{\langle ij \rangle,\sigma}(c_{i\sigma}^{\dagger}c_{j\sigma}+{\rm H.c.})-\sum_{i,\sigma}(\mu_0+\xi_i\mu_{\rm co})c_{i\sigma}^{\dagger}c_{i\sigma} \\ \nonumber
&-\sum_{i}(\Delta_ic_{i\uparrow}^{\dagger}c_{i\downarrow}^{\dagger}+{\rm H.c.})-{i}t_{\rm lc}\sum_{\langle ij \rangle,\sigma}(c_{i\sigma}^{\dagger}c_{j\sigma}-{\rm H.c.}),
\end{align}
where $t$ is the nearest-neighbor hopping energy, $\mu_0$ is the global chemical potential, $\mu_{\rm co}$ is the charge order amplitude, $\xi_i$ is a local variable ($\pm 1$) that generates the tri-hexagonal charge order pattern, shown in Fig.~\ref{fig1}, $\Delta_i$ is the local spin-singlet pairing gap, and the complex nearest-neighbor hopping ${i}t_{\rm lc}$ incorporates the TRS breaking orbital loop current. The Hamiltonian is diagonalized by using the standard unitary transformation of the fermionic fields $c_{i\sigma}\!=\!\sum_n u_{ni}^{\sigma}\gamma_n+v_{ni}^{\sigma *}\gamma_n^{\dagger}$, where $\gamma_n$ is an annihilation operator acting on the $n^{\rm th}$ eigenstate, and $ u_{ni}^{\sigma}$ ($v_{ni}^{\sigma}$) is the corresponding quasiparticle (quasihole) amplitude at site $i$ and spin $\sigma$. The eigenstates are obtained by solving the Bogoliubov-de Gennes equations $\sum_j{\cal H}_{ij}\psi_{nj}\!=\!E_n\psi_{ni}$, subject to the self-consistent gap equation~\cite{supp}
\begin{align}
\Delta({\bf r}_i)=\frac{{\cal U}}{2}\sum_{n} \big[u_{ni}^{\uparrow} v_{ni}^{\downarrow *}-u_{ni}^{\downarrow} v_{ni}^{\uparrow *}\big]\tanh \Big( \frac{E_n}{2k_BT}\Big),
\end{align}
where $\psi_{ni}\!=\![u_{ni}^{\uparrow}, u_{ni}^{\downarrow}, v_{ni}^{\uparrow}, v_{ni}^{\downarrow}]^T$, ${\cal U}$ is the pair-wise attractive potential and $T$ is the temperature. Throughout the presented results, $\mu_0$ is kept at zero which places the Fermi level close to one of the van Hove singularities, $t\!=\!1$ and ${\cal U}\!=\!2$. The relevant energy scale is the maximum pairing gap magnitude, which was found experimentally to be $\Delta\!\approx \! 0.52$~meV~\cite{Chen2021}, is taken here to be the unit for all energies in what follows.

To keep track of the superconducting transition, the global superconducting phase rigidity, determined by the superfluid density, is calculated from the effective Drude weight, given by~\cite{Scalapino_PRL1992,supp}
\begin{align}
n_s=\frac{D_s}{\pi e^2}=-\langle \kappa \rangle+\Lambda({\bf Q}\rightarrow 0,i\omega\rightarrow 0),
\end{align}
where the first term on the right hand side is the diamagnetic response, with the local kinetic energy expressed in terms of the Bogoliubov quasiparticle weights as
\begin{align}
\kappa_{i} =& -t\sum_{\langle j \rangle, n,\sigma}\big[ u_{n i}^{\sigma} u_{n j}^{\sigma*} + {\rm c.c.}\big]f(E_n) \nonumber \\
&+\big[ v_{n i}^{\sigma} v_{n j}^{\sigma*} + {\rm c.c.}\big](1-f(E_n)).
\end{align}
The second term represents the paramagnetic response, obtained by the transverse current-current correlation function 
\begin{align}
\Lambda({\bf Q}\!\rightarrow\!0,i\omega\!\rightarrow\!0)\!&=\!\frac{1}{N}\!\sum_{i,j,n_1,n_2}^{\sigma,\sigma^{\prime}}\!{\cal A}_{n_1n_2}^{i\sigma\sigma^{\prime}} \!\big[ {\cal A}_{n_1n_2}^{j\sigma\sigma^{\prime}*}\!+\!{\cal B}_{n_1n_2}^{j\sigma\sigma^{\prime}} \big]  \nonumber \\
&\times \frac{f(E_{n1})-f(E_{n2})}{E_{n1}-E_{n2}},
\end{align}
where $N$ is the total number of lattice sites and 
\begin{align}
&{\cal A}_{n_1n_2}^{i\sigma\sigma^{\prime}}=2\big[u_{n_1j}^{\sigma^{\prime} *}u_{n_2i}^{\sigma}-u_{n_1i}^{\sigma *}u_{n_2j}^{\sigma^{\prime}} \big],  \nonumber \\
&{\cal B}_{n_1n_2}^{i\sigma\sigma^{\prime}}=2\big[v_{n_1j}^{\sigma^{\prime} *}v_{n_2i}^{\sigma}-v_{n_1i}^{\sigma *}v_{n_2j}^{\sigma^{\prime}} \big].
\end{align}

The local density of states, an observable that can be compared with the scanning tunneling microscopy data, is calculated via
\begin{align}
\rho ({\mathbf{r}_i,E})\!=\!\sum_{n}\big[ |u_{n i}^{\sigma}|^2 \delta(E\!-\!E_n)\!+\!|v_{n i}^{\sigma}|^2 \delta(E\!+\!E_n) \big].
\end{align}
The total density of states $\rho(E)$ is obtained by summing over all lattice sites, and the Fourier transformed local density of states at a momentum $\mathbf{Q}$ is obtained using
\begin{align}
\rho ({\mathbf{Q},E})\!=\!\frac{1}{N}\sum_{i} \cos\!\big(\mathbf{Q} \cdot \mathbf{r}_i\big)\rho({\mathbf{r}_i,E}).
\end{align}
The local and non-local modulations of the density of states are useful to analyze the presence of the particle-hole symmetry and hence, to differentiate the contributions from the normal fluid and the superfluid, as will be evident from the numerical results presented below.

\begin{figure}[t]
\begin{center}
\vspace{-0mm}
\epsfig{file=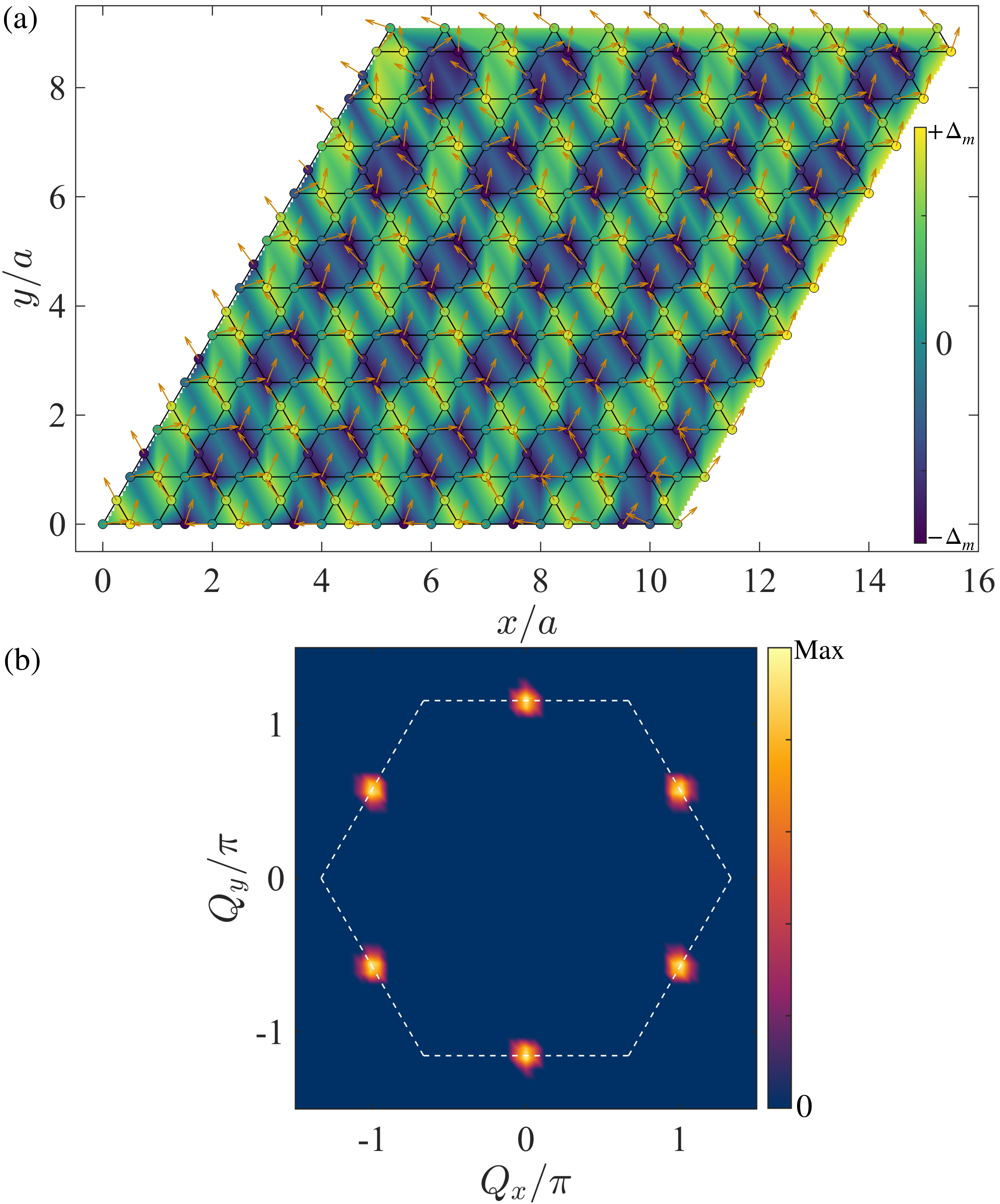,trim=0.0in 0.0in 0.0in 0.0in,clip=false, width=86mm}
\caption{(a) Profile of the pairing gap $\Delta({\bf r}_i)\!=\!\Delta_me^{i\theta_i}$ solution on the considered $10a\times10a$ lattice with periodic boundary conditions---the color scale shows the real part; the arrows show the phase $\theta_i$. (b) Fourier transform of the pair-pair correlation function $C({\mathbf{Q}})$ obtained on a $20a\times20a$ lattice with periodic boundary conditions, showing the three characteristic momenta (six-peak structure), indicative of the CPDW state. The hexagon plotted with dashed lines depicts the Brillouin zone. Parameters used are $\mu_{\rm co}=0.5$ and $t_{\rm lc}=1$.}
\label{fig2}
\vspace{-4mm}
\end{center}
\end{figure}

\textit{The CPDW state.}---The density wave of $s$-wave bosons is envisaged from the pairing gap $\Delta({\bf r}_i)\!=\!\Delta_me^{i\theta_i}$, both real and imaginary parts of which reveal $2a\! \times \!2a$ periodic modulation (real part is shown on the color scale in Fig.~\ref{fig2}(a)). The phase angle $\theta_i$ also shows a periodic structure (arrows in Fig.~\ref{fig2}(a)). This intriguing CPDW state is confirmed further by the Fourier transform of the pair-pair correlation function
\begin{align}
C({\mathbf{Q}})\!=\!\frac{1}{N}\sum_{i,j} \langle \Delta(\mathbf{r}_i) \Delta(\mathbf{r}_j) \rangle e^{-i\mathbf{Q} \cdot \mathbf{r}_{ij}},
\end{align}
which shows three characteristic momenta (see Fig.~\ref{fig2}(b)), given by ($\pm\pi$, $\pi/\sqrt{3}$) and (0, 2$\pi/\sqrt{3}$). It is confirmed that the CPDW state is generated in the superconducting state due to the interplay between $s$-wave onsite pairing, charge order and the TRS breaking orbital loop current. Three types of such triple-$\mathbf{Q}$ correlation, at different momenta, have been observed in the experiments~\cite{Zhao2021,Chen2021}, implying that the density waves are cascaded between normal fluid and superfluid \textit{i.e.} the CPDW can also induce subsequent charge orders.
%---------------------------------------------
%---------------------------------------------

%---------------------------------------------
\begin{figure}[b]
\begin{center}
\vspace{0mm}
\epsfig{file=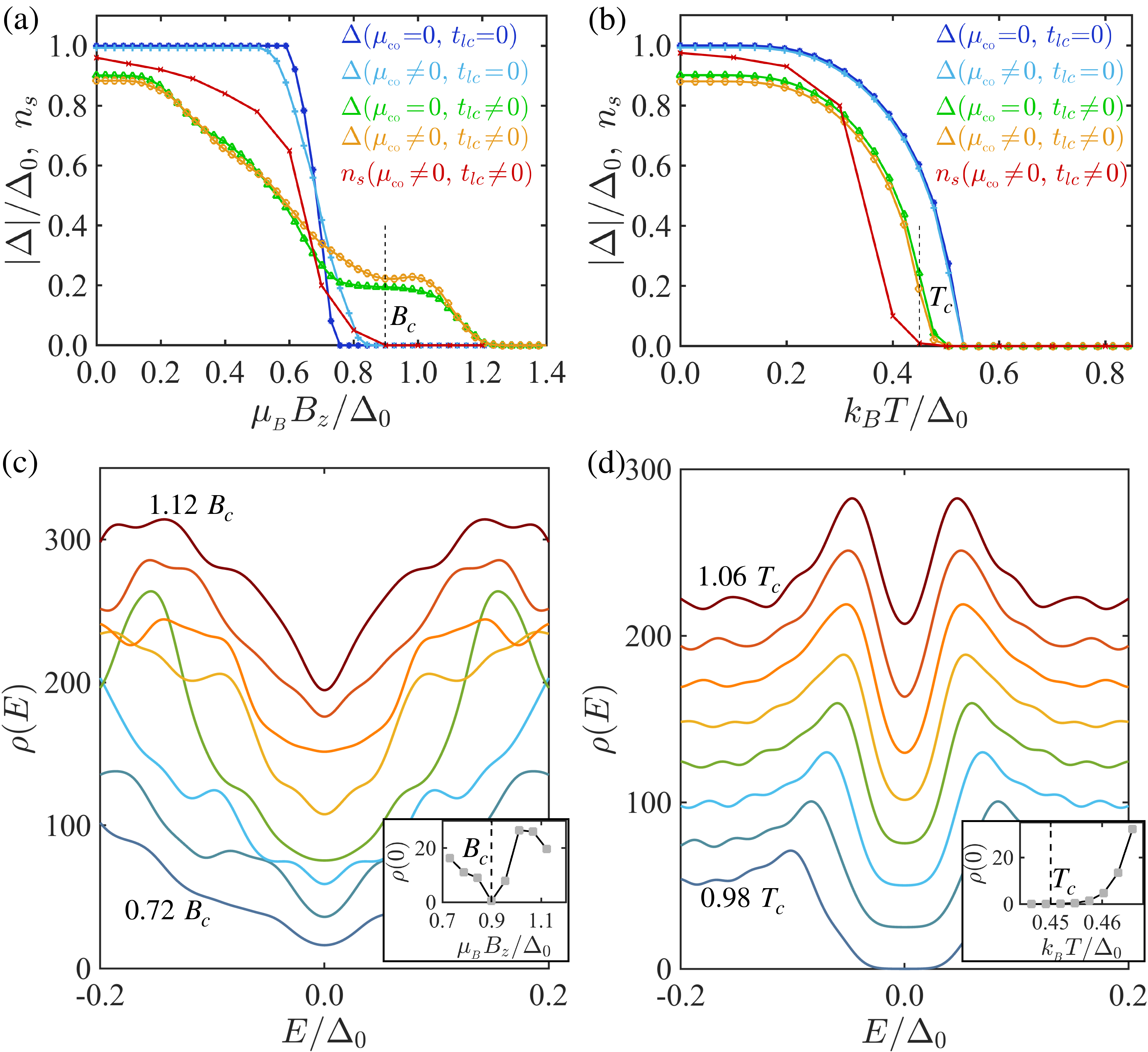,trim=0.0in 0.0in 0.0in 0.0in,clip=false, width=86mm}
\caption{(a), (b) Variation of the average gap magnitude $|\Delta|$ (without and with the charge order and the loop current) and superfluid density $n_s$ with magnetic field $B_z$ and temperature $T$. (c), (d) Density of states $\rho(E)$ for different $B_z$ and $T$ near the critical values $B_c$ and $T_c$, determined by vanishing $n_s$. Insets in (c), (d) show the density of states at zero energy $\rho(0)$ as a function of $B_z$ and $T$, respectively. The results were obtained on a $20a\times20a$ lattice with periodic boundary conditions. All other parameters are the same as in Fig.~2. A constant offset has been added to the vertical axis for each curve in (c) and (d) for clarity.}
\label{fig3}
\vspace{-5mm}
\end{center}
\end{figure}
%---------------------------------------------
\textit{Pseudogap.}---The coexistence of charge order and superconductivity in AV$_3$Sb$_5$ over a large parameter regime raised the natural question whether there is a cooperation between the two commonly-known competing orders~\cite{Yu_NComm2021}. The present analysis shows that the superconducting gap is suppressed in the presence of the charge order and the orbital loop current at zero temperature and zero magnetic field. However, the average pairing gap $|\Delta|$ vanishes at a magnetic field and a temperature, larger than the critical values $B_c$ and $T_c$, determined by a vanishing superfluid density $n_s$ (Fig.~\ref{fig3}(a)-(b)). The magnetic field of amplitude $B_z$ was incorporated by the Hamiltonian ${\cal H}_{_Z}\!=\!-\mu_{_B}B_z\sum_{i,\sigma,\sigma^{\prime}}{\boldsymbol \sigma}_{\sigma \sigma^{\prime}}^z c_{i\sigma}^{\dagger}c_{i\sigma^{\prime}}$ which describes the Zeeman exchange coupling. Remarkably, the critical field at which $|\Delta|$ drops to zero is enhanced by more than 20$\%$ above $B_c$ in the presence of the charge order and the loop current. 
%---------------------------------------------
\begin{figure*}[t]
\begin{center}
\vspace{0mm}
\epsfig{file=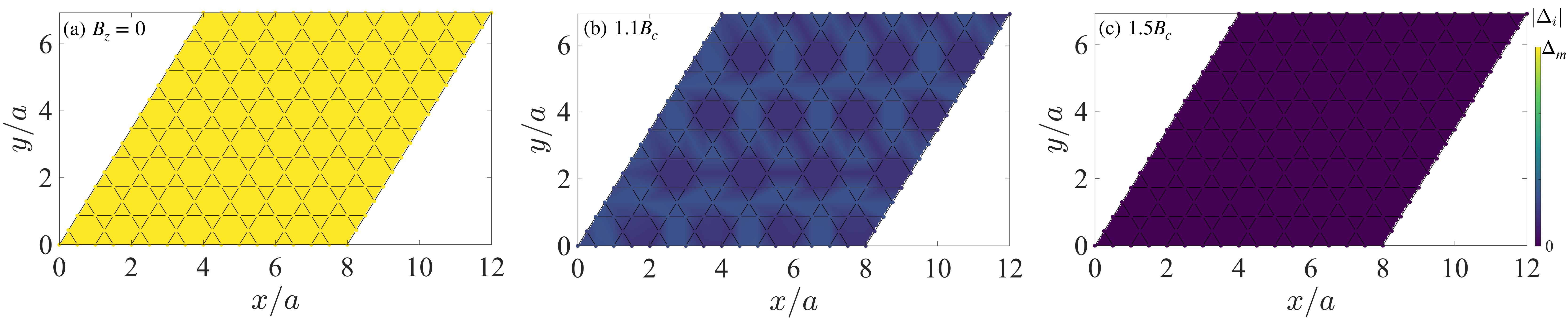,trim=0.0in 0.0in 0.0in 0.0in,clip=false, width=180mm}
\caption{Profile of the absolute value of the pairing gap $|\Delta({\bf r}_i)|$, self-consistently obtained on a $8a\times8a$ lattice with periodic boundary conditions, at different magnetic fields across the superconducting transition: (a) $B_z\!=\!0$, (b) $B_z\!=\!1.1B_c$, and (c) $B_z\!=\!1.5B_c$. In the vicinity of the superconducting transition (the case of plot (b)), the pairing gap is locally suppressed in the charge-ordered clusters, creating a granular phase having superconducting patches separated by non-superconducting regions. All other parameters are the same as in Fig.~2.}
\label{fig4}
\vspace{-2mm}
\end{center}
\end{figure*}
%---------------------------------------------
%---------------------------------------------
\begin{figure}[b]
\begin{center}
\vspace{0mm}
\epsfig{file=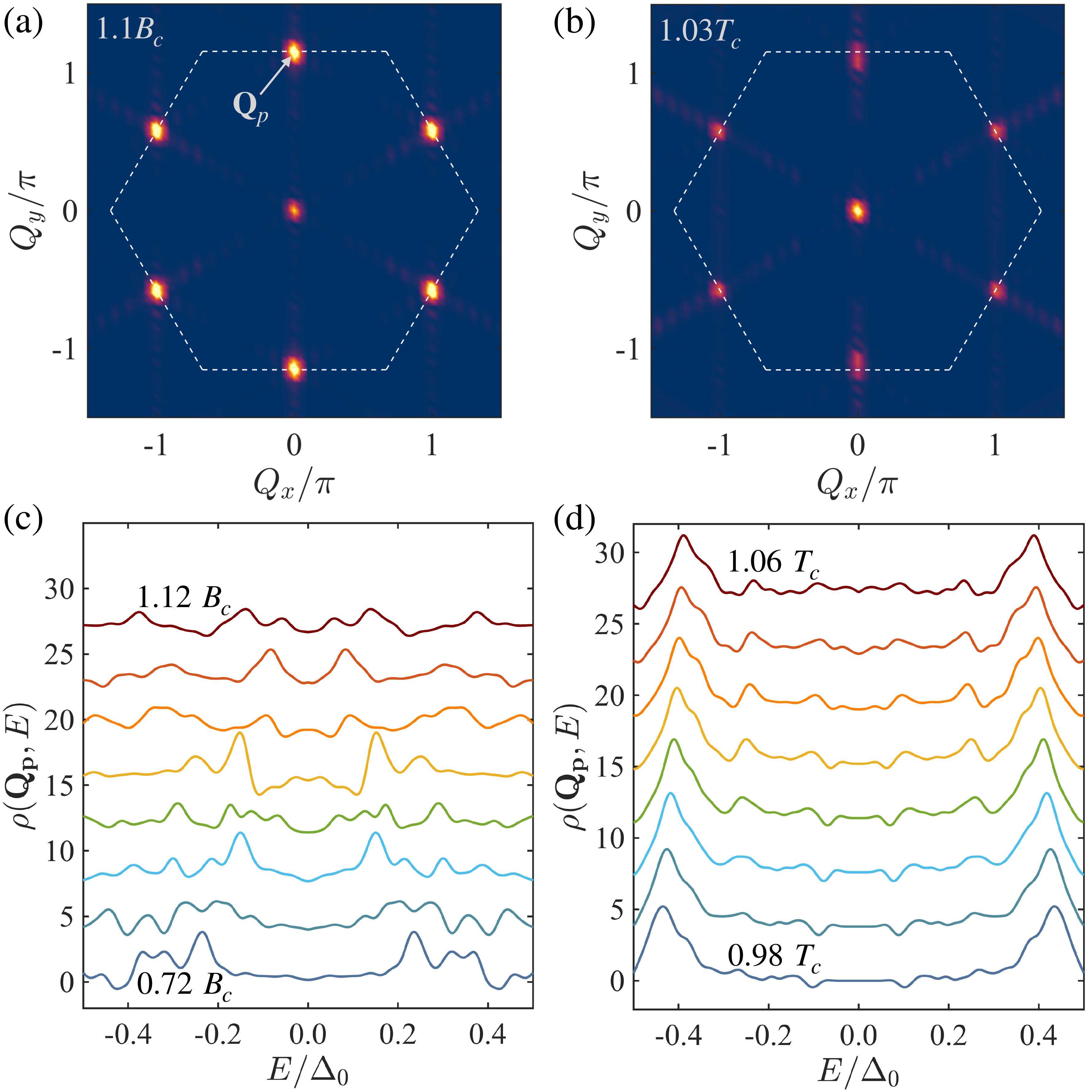,trim=0.0in 0.0in 0.0in 0.0in,clip=false, width=86mm}
\caption{Fourier transformed pair-pair correlation function $C({\mathbf{Q}})$ at (a) magnetic field $B_z\!=\!1.1B_c$, temperature $T\!=\!0$, and (b) $B_z\!=\!0$, $T\!=\!1.03T_c$, indicating the presence of the CPDW correlation above the superconducting transition. (c), (d) Fourier-transformed local density of states $\rho ({\mathbf{Q}_p,E})$ at a characteristic CPDW momentum $\mathbf{Q}_p$, shown in (a), at different values of $B$ and $T$ across the superconducting transition, revealing particle-hole symmetric quasiparticle states with a non-zero weight at $E\!=\!0$. The results were obtained on a $20a\times20a$ lattice with periodic boundary conditions. All other parameters are the same as in Fig.~2. A constant offset has been added to the vertical axis for each curve in (c) and (d) for clarity.}
\label{fig5}
\vspace{-4mm}
\end{center}
\end{figure}
%---------------------------------------------
It is known that a conventional spin-singlet superconductor, with a gap around the Fermi level, has a vanishing paramagnetic susceptibility at $T\!=\!0$, and hence it cannot lower its Free energy indefinitely by spin-polarizing the quasiparticle states in the presence of a Zeeman magnetic field. Consequently, when the Zeeman energy gain is comparable to the superconducting condensation energy, given by $\mu_{_B}B_{c0}\!=\!\Delta_0/\sqrt{2}\!\approx\!0.7\Delta_0$, known as the Chandrasekhar-Clogston limit~\cite{Chandrasekhar1962,Clogston1962}, there is a transition to the normal state. An exception to this stringent condition occurs in the case of Fulde-Ferrell-Larkin-Ovchinikov type finite-momentum condensates~\cite{Fulde_Ferrell,Larkin_Ovchinnikov} and in thin superconducting films with a large spin-orbit coupling~\cite{Bruno_PRB1973}. The enhancement in the critical field for vanishing $|\Delta|$ in the kagom\'e lattice with a charge order and orbital loop current indicates the formation of an unconventional state above and in the vicinity of the superconducting transition. However, it can be attributed to the non-zero density of states $\rho(E)$ within the superconducting gap (Fig.~\ref{fig3}(c)) in the pseudogap state. At the critical field $B_c$ for vanishing $n_s$, $|\Delta|$ exhibits a dip while $\rho(E\!=\!0)$ shows a zero weight (inset in (Fig.~\ref{fig3}(c))), attesting the appearance of an unconventional state immediately above $B_c$. The enhancement of the critical magnetic field and the appearance of the pseudogap in the CPDW state suggest that there is a correlation among these phenomena. The temperature driven transition to the normal state also reveals a similar pseudogap (Fig.~\ref{fig3}(d)), though the variation of $\rho(E\!=\!0)$ with $T$ is rather monotonic. Moreover, the V-shaped density of states and the multiple coherence peaks around the gap show similarities with those observed in the tunneling spectra~\cite{Chen2021,Zhao2021,Xu_PRl2021}.

\textit{CPDW correlation above superconducting transition.}---To gain insights into the origin of the pseudogap, the profile of the absolute value of the pairing gap at different fields across the superconducting transition was looked at in Fig.~\ref{fig4}(a),(b),(c). It is found that above and in the vicinity of the critical magnetic field, the pairing gap is locally suppressed in charge-ordered clusters, stabilizing an emergent granular superconducting state in which the pairing gap survives locally even though the long-range superconducting phase coherence vanishes. The magnetization profile also reveals periodic modulation, and the interplay of magnetization and superconducting gap prevails in the entire granular superconducting phase~\cite{supp}. The fourier-transformed pair-pair correlation function $C({\mathbf{Q}})$ and the fourier-transformed local density of states $\rho ({\mathbf{Q}_p,E})$ at $\mathbf{Q}_p$, one of the three characteristic momenta for the CPDW, were also investigated (Fig.~\ref{fig5}(a)-(d)) across the superconducting transition. Interestingly, the CPDW correlation survives above the critical magnetic field $B_c$ and critical temperature $T_c$. The observable $\rho ({\mathbf{Q}_p,E})$ is particle-hole symmetric \textit{i.e.} it is symmetric when $E\!\rightarrow\!-E$, above and in the vicinity of $B_c$ and $T_c$, ruling out other possible mechanisms of the pseudogap such as charge order of electronic states or modulation due to electron scattering from a periodic potential. From these findings, it can be argued that the modulations of the density of states in the pseudogap, observed in the experiments, are a consequence of the CPDW of $s$-wave Cooper pairs without a global phase coherence.

\textit{Discussion and conclusion.}---The superconducting state can be influenced by multiple properties of the compounds such as the TRS breaking loop current, the rotational symmetry-breaking nematic order, fermi surface nesting, Coulomb interactions and sublattice interference~\cite{Xiang2021,Jin_PRL2021,Nie2022,Zhou2022,Wu_PRL2022}. Despite the complex nature of the pairing mechanism, there are growing experimental evidences in support of spin-singlet $s$-wave pairing such as the absence of a nodal state while transitioning from an anisotropic full-gap state to an isotropic full-gap state driven by impurity concentration~\cite{Roppongi2023}, and the appearance of a prominent Hebel-Slichter coherence peak immediately below $T_c$~\cite{Mu_ChPhysLett2021}. The proposed CPDW of $s$-wave Cooper pairs, therefore, provides a natural explanation for many paradoxical experimental observations, including the pseudogap in the tunneling spectra. 

To summarize, it is shown that a CPDW state of $2a\! \times \!2a$ periodicity emerges spontaneously in the kagom\'e lattice due to the interplay of onsite spin-singlet superconductivity with a charge order of the same periodicity and an orbital loop current. The CPDW correlation survives beyond the superconducting transition in granular regions without global superconducting phase coherence, producing a V-shaped particle-hole symmetric density of states and pseudogap, which are otherwise indicative of a nodal unconventional pairing symmetry.

\textit{Acknowledgements.}---The work was supported by Science and Engineering Research Board, India (Research Grant No. SRG/2023/001188). Numerical calculations were performed at the computing resources of PARAM Ganga at the Indian Institute of Technology Roorkee, provided by National Supercomputing Mission, implemented by C-DAC, and supported by the Ministry of Electronics and Information Technology and Department of Science and Technology, Government of India.

\vspace{-1mm}
%\bibliography{Ref}


%%%%%%%%%% Merge with supplemental materials %%%%%%%%%%
\newpage
\pagebreak
%\newpage
\clearpage
\widetext
\begin{center}
\textbf{{Supplemental Materials for "Chiral pair density wave as a precursor of the pseudogap in kagom\'e superconductors"}}
%\vspace{1em}\\
%{Narayan Mohanta}\\
%\it{Department of Physics, Indian Institute of Technology Roorkee, Roorkee 247667, India}

\end{center}
%%%%%%%%%% Merge with supplemental materials %%%%%%%%%%
%%%%%%%%%% Prefix a "S" to all equations, figures, tables and reset the counter %%%%%%%%%%
\setcounter{equation}{0}
\setcounter{figure}{0}
\setcounter{table}{0}
\setcounter{page}{1}
\makeatletter
\renewcommand{\theequation}{E\arabic{equation}}
\renewcommand{\thefigure}{S\arabic{figure}}
\renewcommand{\bibnumfmt}[1]{[R#1]}
\renewcommand{\citenumfont}[1]{R#1}
%%%%%%%%%% Prefix a "S" to all equations, figures, tables and reset the counter %%%%%%%%%%

%\vspace{5mm}

%\noindent Contents of this supplemental material:\\
%\\1. Robustness of the CPDW state\\
%2. Self-consistent gap equation\\
%3. Calculation of superfluid density\\
%4. Magnetization in the emergent granular state\\
%5. Effect of loop current in the density of states\\

\begin{figure}[b]
\begin{center}
\vspace{-0mm}
\epsfig{file=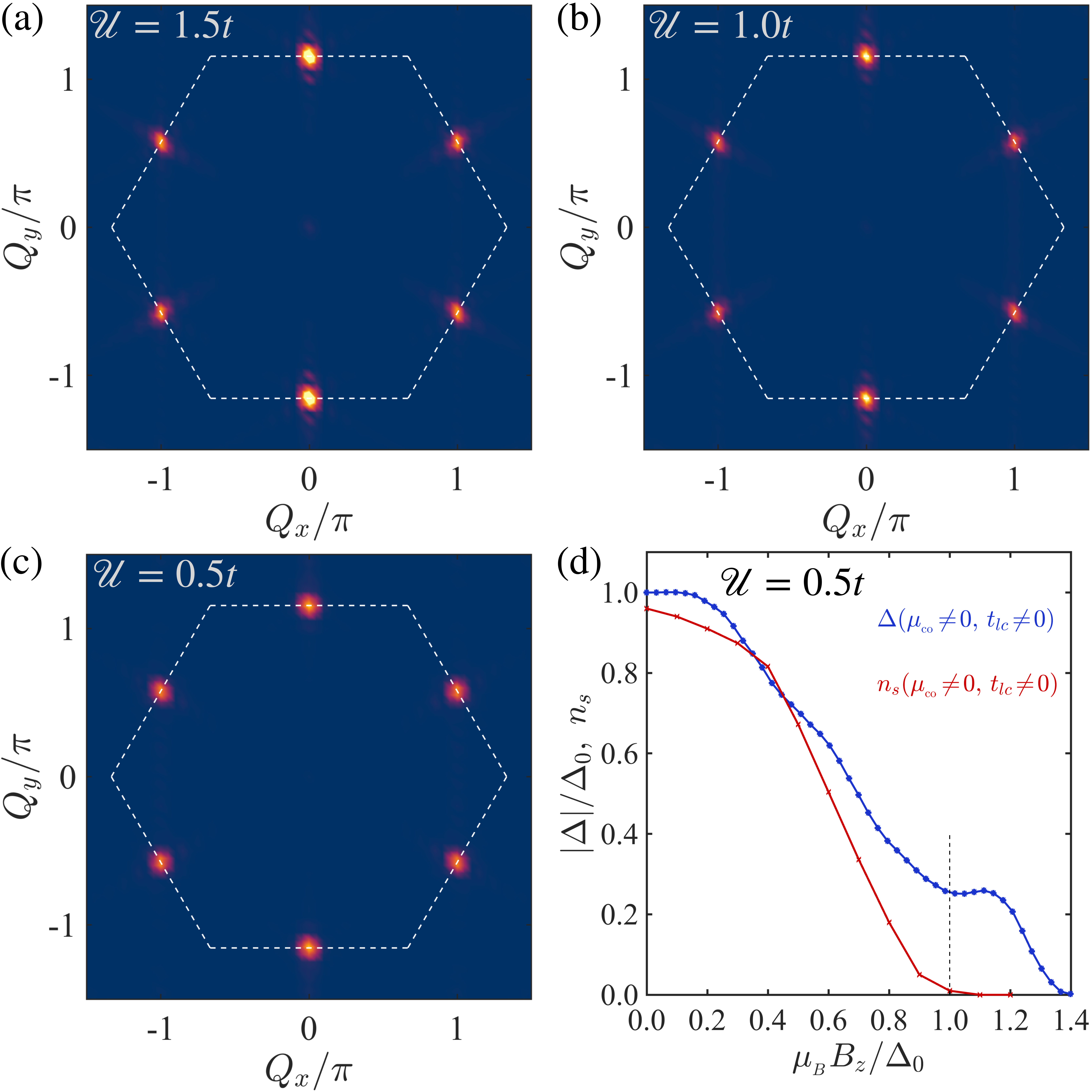,trim=0.0in 0.0in 0.0in 0.0in,clip=false, width=87mm}
\caption{(a)-(c) Fourier transformed pair-pair correlation function $C({\mathbf{Q}})$ at different values of ${\cal U}$: (a) ${\cal U}\!=\!1.5t$, (b) ${\cal U}\!=\!1.0t$, and (c) ${\cal U}\!=\!0.5t$. (d) Normalized pairing gap magnitude $|\Delta|/\Delta_0$ and the superfluid density $n_s$ as a function of the magnetic field $B_z$. The results were obtained on a $20a\times20a$ lattice with periodic boundary conditions. All other parameters are the same as in Fig.~1 of the main text.}
\label{figS1}
\vspace{-4mm}
\end{center}
\end{figure}

\begin{figure}[t]
\begin{center}
\vspace{-0mm}
\epsfig{file=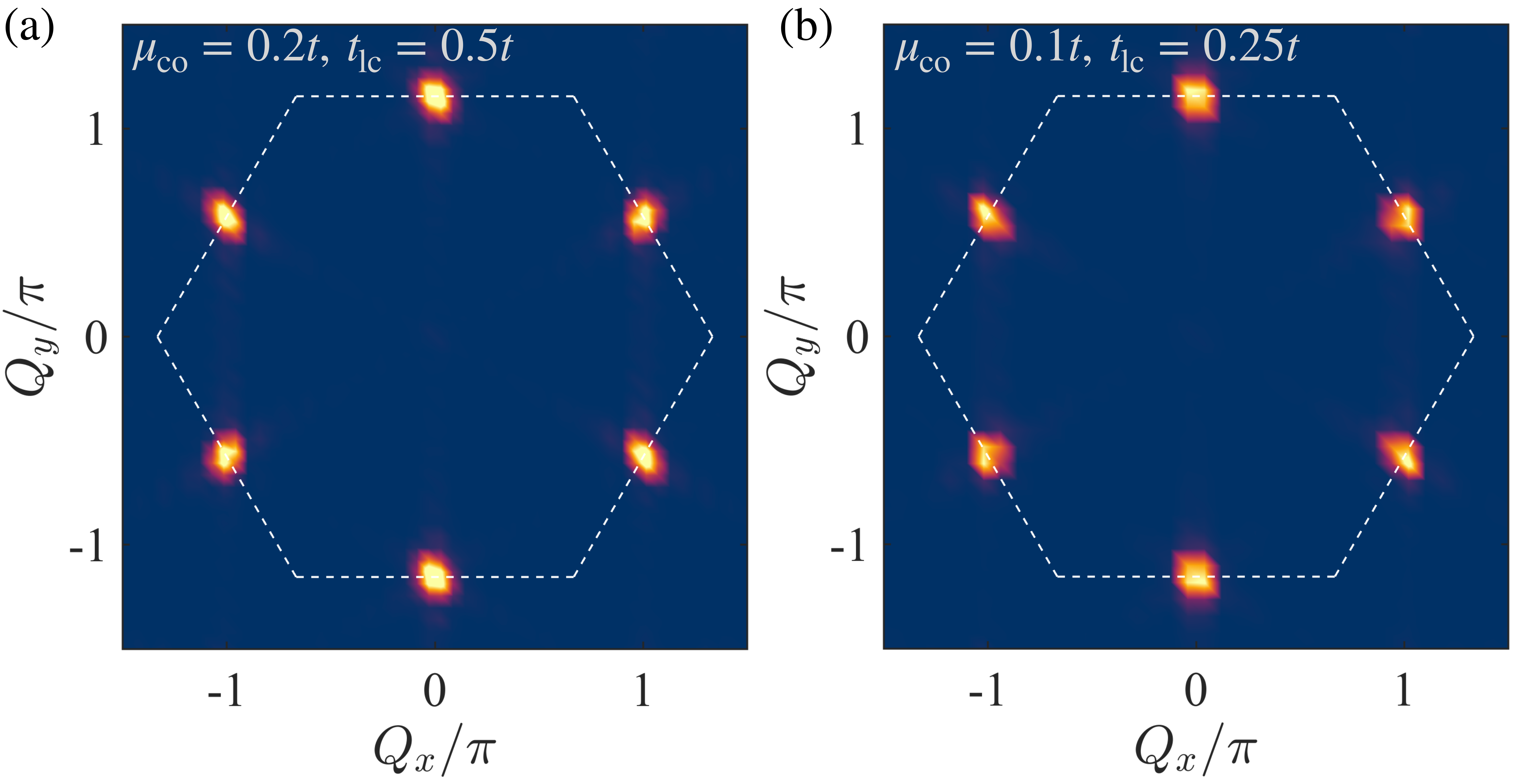,trim=0.0in 0.0in 0.0in 0.0in,clip=false, width=87mm}
\caption{Fourier transformed pair-pair correlation function $C({\mathbf{Q}})$ at different values of the charge-order strength $\mu_{\rm co}$ and loop current parameter $t_{\rm lc}$: (a) $\mu_{\rm co}\!=\!0.2t$, $t_{\rm lc}\!=\!0.5t$, and (b) $\mu_{\rm co}\!=\!0.1t$, $t_{\rm lc}\!=\!0.25t$. The results were obtained on a $20a\times20a$ lattice with periodic boundary conditions. All other parameters are the same as in Fig.~1 of the main manuscript.}
\label{figS2}
\vspace{-4mm}
\end{center}
\end{figure}

\noindent {\bf \\1. Robustness of the CPDW state}\\
\noindent The CPDW state, discussed in the main text, appears also for smaller values of the attractive pair-wise interaction strength ${\cal U}$, as shown in Fig.~\ref{figS1}(a), (b), (c). The discrepancy between the pairing amplitude and the superfluid density exists at smaller values of ${\cal U}$, as shown in Fig.~\ref{figS1}(d). Similarly, different sets of the charge-order strength $\mu_{\rm co}$ and loop current parameter $t_{\rm lc}$ also give rise to the CPDW state and the pseudogap behavior near the superconducting transition. Fig.~\ref{figS2} shows the fourier transformed pair-pair correlation function $C({\mathbf{Q}})$ for two sets of $\mu_{\rm co}$ and $t_{\rm lc}$, showing the C3-symmetric CPDW correlations.

%---------------------------------------------
%---------------------------------------------

%---------------------------------------------
%---------------------------------------------

%---------------------------------------------
\begin{figure}[t]
\begin{center}
\vspace{-0mm}
\epsfig{file=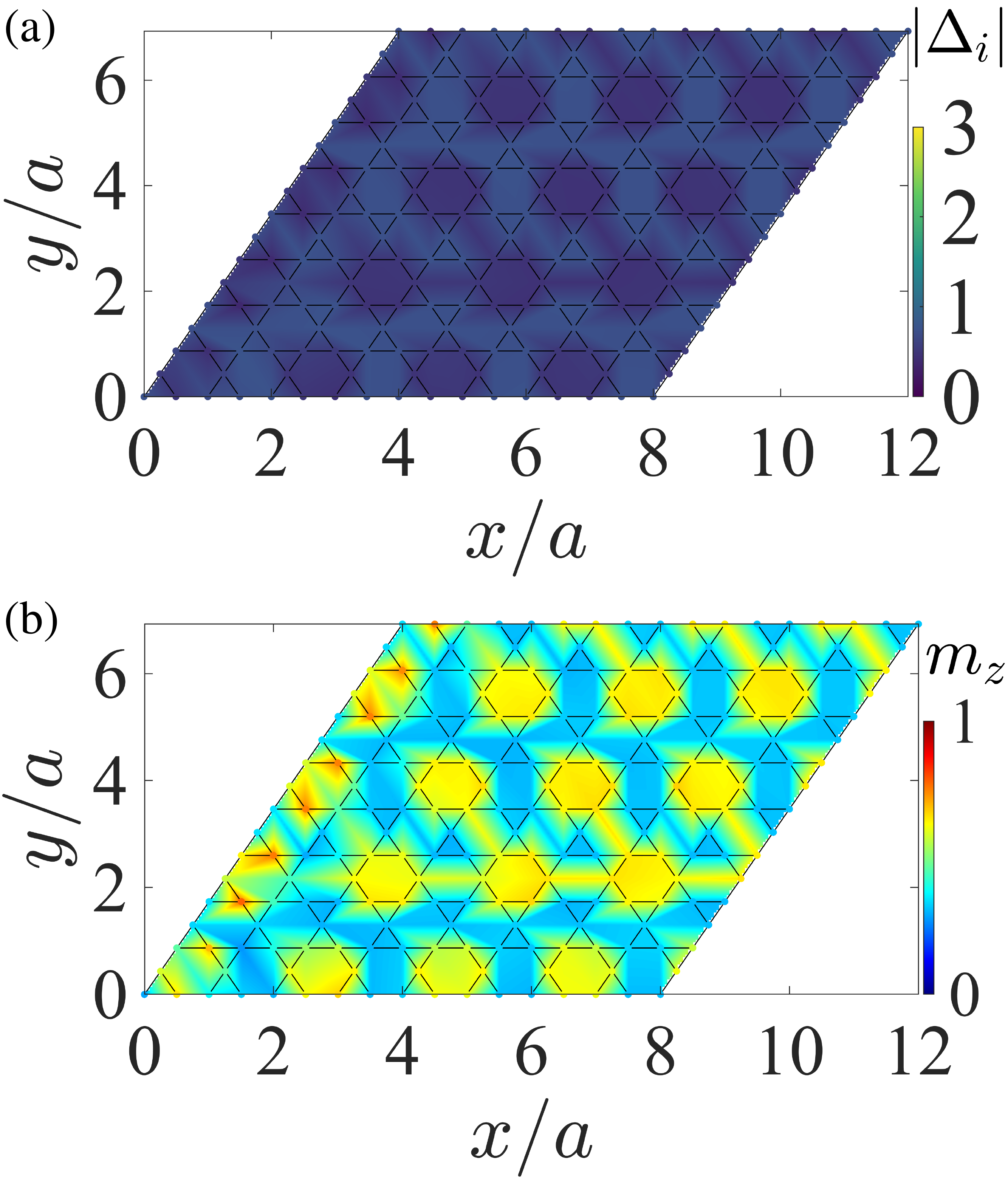,trim=0.0in 0.0in 0.0in 0.0in,clip=false, width=85mm}
\caption{Profile of the absolute value of the pairing gap (top panel) and $z$ component of magnetization (bottom panel) at $B\!=\!1.1B_c$. All other parameters are the same as in Fig.4 of the main text.}
\label{figS3}
\vspace{-4mm}
\end{center}
\end{figure}
%---------------------------------------------
\noindent {\bf \\2. Self-consistent gap equation}\\
\noindent The superconducting order parameter, the local pairing gap $\Delta({\bf r}_i)\!=\!-{\cal U}\langle c_{i\uparrow}c_{i\downarrow} \rangle$, can be re-written using the Bogoliubov-Valatin transformation $c_{i\sigma}\!=\!\sum_n u_{ni}^{\sigma}\gamma_n+v_{ni}^{\sigma *}\gamma_n^{\dagger}$, where $\gamma_n$ is fermionic annihilation operator acting on the $n^{\rm th}$ eigenstate and $ u_{ni}^{\sigma}$ ($v_{ni}^{\sigma}$) is the corresponding quasiparticle (quasihole) amplitude at site $i$ and spin $\sigma$, as\\
 \begin{align}
\Delta({\bf r}_i)=-{\cal U} \sum_{n} \big[u_{ni}^{\uparrow} v_{ni}^{\downarrow*}(1-f(E_n))+u_{ni}^{\downarrow} v_{ni}^{\uparrow*}f(E_n) \big]
\end{align}
where $f(E_n)$ is the fermi function corresponding to the $n^{\rm th}$ eigen-energy $E_n$. Using the relation $f(x)=\frac{1}{2}-\frac{1}{2}\tanh{(\frac{x}{2})}$, the above equation can be expressed as
\begin{align}
\Delta({\bf r}_i)&=\frac{{\cal U}}{2}\sum_{n} \big[u_{ni}^{\uparrow} v_{ni}^{\downarrow *}-u_{ni}^{\downarrow} v_{ni}^{\uparrow *}\big]\tanh \Big( \frac{E_n}{2k_BT}\Big),
\end{align}
where $k_B$ is the Boltzmann constant and $T$ is the temperature. In the above equation, the energy-independent terms were dropped because the contribution from those terms will vanish. The above equation is the self-consistent gap equation used in the calculations. The self-consistency iteration is performed until convergence is achieved at each lattice sites.

\noindent {\bf \\3. Calculation of superfluid density}\\
The superfluid density is given by the effective Drude weight, derived earlier in Ref.~[\onlinecite{Scalapino_PRL1992}]
\begin{align}
n_s=\frac{D_s}{\pi e^2}=-\langle \kappa \rangle+\Lambda({\bf Q}\rightarrow 0,i\omega\rightarrow 0),
\end{align}
where the first term on the right hand side is the diamagnetic response and the second term is the paramagnetic response. The diamagnetic term, associated with the local kinetic energy, can be expressed in terms of the Bogoliubov-de Gennes (BdG) quasiparticle weights as
\begin{align}
\kappa_{i} =& -t\sum_{\langle j \rangle, n,\sigma}\big[ u_{n i}^{\sigma} u_{n j}^{\sigma*} + c.c.\big]f(E_n) \nonumber \\
&+\big[ v_{n i}^{\sigma} v_{n j}^{\sigma*} + c.c.\big](1-f(E_n)).
\end{align}
The paramagnetic response is given by the transverse current-current correlation function
\begin{align}
\Lambda({\bf Q},\omega_n)\!&=\!\frac{1}{N} \int_0^{1/T}e^{i\omega_n \tau} \langle j_x^p({\bf Q},\tau) j_x^p(-{\bf Q},0) \rangle~d\tau,
\end{align}
with $\omega_n\!=\!2\pi nT$ and the paramagnetic current $j_x^p({\bf Q})$ is given by 
\begin{align}
j_x^p({\bf Q})=it\sum_{i,\sigma}(c_{i+\hat{x},\sigma}^{\dagger}c_{i,\sigma}-c_{i,\sigma}^{\dagger}c_{i+\hat{x},\sigma})e^{-{\bf Q}\cdot {\bf r}_i}.
\end{align}

The uniform (i.e. {\bf Q}={\bf 0}) paramagnetic response can be obtained~\cite{Scalapino_PRL1992}, in terms of the BdG quasiparticle weights as 
\begin{align}
\Lambda({\bf Q}\!\rightarrow\!0,i\omega\!\rightarrow\!0)\!&=\!\frac{1}{N}\!\sum_{i,j,n_1,n_2}^{\sigma,\sigma^{\prime}}\!{\cal A}_{n_1n_2}^{i\sigma\sigma^{\prime}} \!\big[ {\cal A}_{n_1n_2}^{j\sigma\sigma^{\prime}*}\!+\!{\cal B}_{n_1n_2}^{j\sigma\sigma^{\prime}} \big]  \nonumber \\
&\times \frac{f(E_{n1})-f(E_{n2})}{E_{n1}-E_{n2}},
\end{align}
where $N$ is the total number of lattice sites and 
\begin{align}
&{\cal A}_{n_1n_2}^{i\sigma\sigma^{\prime}}=2\big[u_{n_1j}^{\sigma^{\prime} *}u_{n_2i}^{\sigma}-u_{n_1i}^{\sigma *}u_{n_2j}^{\sigma^{\prime}} \big],  \nonumber \\
&{\cal B}_{n_1n_2}^{i\sigma\sigma^{\prime}}=2\big[v_{n_1j}^{\sigma^{\prime} *}v_{n_2i}^{\sigma}-v_{n_1i}^{\sigma *}v_{n_2j}^{\sigma^{\prime}} \big].
\end{align}\\
%---------------------------------------------
\begin{figure*}[t]
\begin{center}
\vspace{-0mm}
\epsfig{file=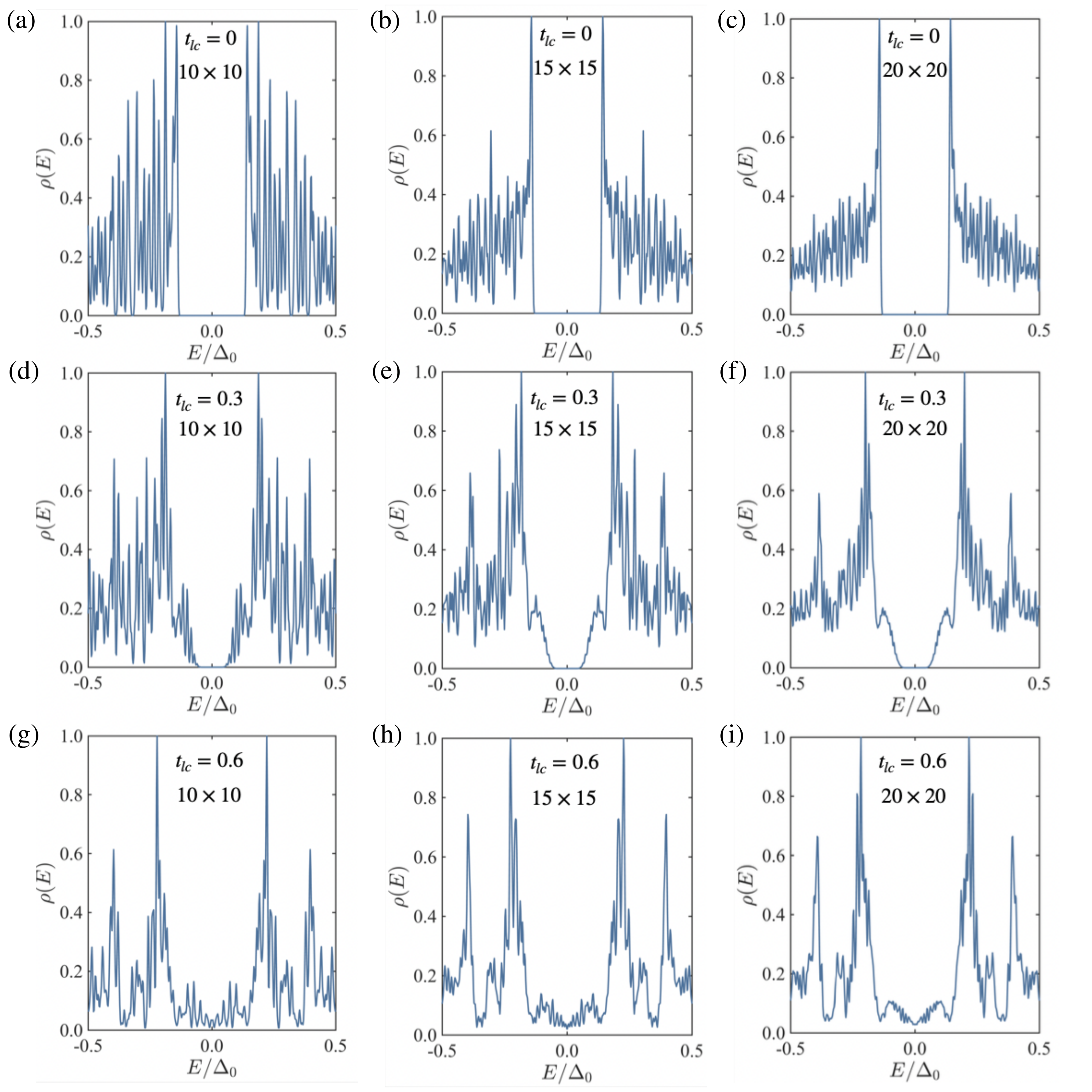,trim=0.0in 0.0in 0.0in 0.0in,clip=false, width=160mm}
\caption{Density of states of a kagome lattice with periodic boundary conditions for different lattice sizes (different columns) and different values of the loop current amplitude $t_{lc}$ (different rows), without chemical potential modulation and at a magnetic field $B\!=\!2t$. }
\label{figS4}
\vspace{-4mm}
\end{center}
\end{figure*}
%---------------------------------------------

\noindent {\bf \\4. Magnetization in the emergent granular state}\\
Near the critical magnetic field $B_c$ for the superconducting transition (as dictated by the vanishing superfluid density), the system goes through an emergent granular phase in which the superconducting gap magnitude is suppressed in the charge-ordered clusters, as shown in Fig.~\ref{figS3}(a) (and Fig.4 in the main text). The magnetization profile, obtained via $m_{zi}=c_{i\uparrow}^{\dagger}c_{i\uparrow}-c_{i\downarrow}^{\dagger}c_{i\downarrow}$, shown in Fig.~\ref{figS3}(b), also reveals a periodic modulation. The magnetization is enhanced slightly in the regions in which the pairing amplitude is decreased. This interplay of magnetization and superconducting order parameter exists in the entire granular superconducting phase, without any long-range superconducting phase coherence.

\noindent {\bf \\5. Effect of loop current in the density of states}\\
To explore the role of the loop current in formation of the pseudogap near the superconducting transition and to understand the finite size effect, in Fig.~\ref{figS4} the density of states has been presented for loop current magnitude $t_{lc}$ and different lattice sizes. The results show that, despite the little noise in the density of state originating due to finite size effect, it is evident that quasiparticle states build up within the superconducting gap due to the loop current.

\vspace{-1mm}

\end{document}